\newcommand{\beq} {\begin{eqnarray}}
\newcommand{\eeq} {\end{eqnarray}}

\newcommand{\half}{\frac{1}{2}}

\documentclass[12pt, preprint,preprintnumbers,amsfonts,aps,nofootinbib]{revtex4}
\usepackage{graphicx}
\begin{document}

\title{Modulino Dark Matter and the INTEGRAL 511 keV Line}

\author{Nathaniel J. Craig}
\affiliation{Department of Physics, Stanford University, Stanford, CA 94305}

\author{Stuart Raby}
\affiliation{Department of Physics, The Ohio State University, Columbus, OH 43210}

\preprint{SU-ITP- 09/37}
\preprint{OHSTPY-HEP-T-09-002}

\begin{abstract}
In this paper we present a simple extension of the minimal
supersymmetric  standard model [MSSM] which ``naturally" produces
the INTEGRAL photon signal.   The model can be embedded in an SU(5)
grand unified theory [GUT] with gauge mediated SUSY breaking.   The
new ingredients are the addition of several MSSM singlets/moduli.
While the masses of the singlets are at the weak scale, their mass
splittings are suppressed by chiral symmetry breaking and naturally
lie around an MeV.  The decay of the heavier modulino to the lighter
one with the associated production of electron - positron pairs
explains the INTEGRAL signal.  Finally, the detection of diffuse
gamma rays from internal bremsstrahlung in the galactic halo would
be a suggestive indication of dark matter decays associated with the
511 keV line, and is an unambiguous additional prediction of this
model.

\end{abstract}

\maketitle

\section{Introduction}

The nature of dark matter remains one of the most pressing problems
of cosmology and particle physics alike. Although many conventional
dark matter candidates may be probed by particle colliders and
direct-detection experiments, the most suggestive signals --
particularly for dark matter candidates with extremely weak
interactions -- may come from indirect measurements. Indeed, the
recent spate of astrophysical anomalies suggests that the most
telling signatures of dark matter may come from the sky  (for a
comprehensive review, see \cite{Hooper:2009zm} and references
therein).

Although recent astrophysical anomalies measured by PAMELA, ATIC,
FERMI, and HESS have drawn attention to dark matter physics above
100 GeV, there remain decisive signatures from much lower energies.
The 511 keV line from the galactic center, as measured by the SPI
spectrometer of the INTEGRAL satellite \cite{Knodlseder:2003sv,
Jean:2003ci, Knodlseder:2005yq, Weidenspointner:2008zz}, corresponds
to a measured photon flux \beq \Phi_{\text{exp}} = \left(9.35 \pm
0.54 \right) \times 10^{-4} \text{ph cm}^{-2} \text{s}^{-1} \eeq
with a large component from the galactic bulge. The emission is
approximately spherically symmetric with a $\sim 6 ^\circ$ FWHM,
potentially with additional unresolved components along the plane of
the galaxy. The 511 keV line (with a 3 keV line width set by
intrumental resolution) is dominated by $e^+ + e^{-}$ annihilations
via positronium \cite{Kinzer:2001ba}. A variety of Standard Model
processes have been suggested to explain the positron excess,
including, e.g., production within Type Ia supernovae and low-mass
X-ray binaries \cite{Knodlseder:2005yq}. These explanations require
some degree of speculative astrophysics in order to explain both the
nature and intensity of the signal, and to date it remains unclear
whether astrophysical explanations may suffice
\cite{Weidenspointner:2008zz}.

Thus far suggestions for origins of the 511 keV signal from physics
beyond the  Standard Model have largely focused around the
annihilations or decays of light, MeV-scale dark matter (see,
e.g.,\cite{Boehm:2003bt,Picciotto:2004rp,Hooper:2004qf,Oaknin:2004mn,Ferrer:2005xva,Bird:2004ts,Serpico:2004nm,Foot:2004kd,Kawasaki:2005xj,Kasuya:2005ay,Rasera:2005sa,Takahashi:2005kp,Kasuya:2006kj,Fayet:2006sa,Cumberbatch:2006bj,Zhitnitsky:2006tu,Frere:2006hp, Conlon:2007gk}).
These models confront a common challenge: additional
MeV-scale sectors are not particularly natural and, if constituting
cold dark matter, require no small degree of engineering to obtain
acceptable relic abundance and phenomenology. A perhaps more natural
scenario might involve TeV-scale dark matter candidates with
MeV-scale splittings between components; the decays or annihilations
of such dark matter may account for the 511 keV signal from
weak-scale hidden sectors \cite{Finkbeiner:2007kk, Pospelov:2007xh,
Cembranos:2008bw, ArkaniHamed:2008qn, Chen:2009dm, Chen:2009ab}. Nonetheless, the origin of the MeV-scale
splitting among dark matter components itself requires further
explanation.

In this paper we present a simple extension of the minimal
supersymmetric  standard model [MSSM] which ``naturally" produces
the INTEGRAL photon signal.   The model can be embedded in an SU(5)
grand unified theory [GUT] with gauge mediated SUSY breaking.   The
new ingredients are the addition of several MSSM singlets. While the
masses of the singlets are at the weak scale, their mass splittings
are suppressed by chiral symmetry breaking and naturally lie around
an MeV.

\section{Model}

Consider a pair of gauge singlet chiral superfields $\Phi_1,
\Phi_2,$ coupled to the Higgs fields $H_u, H_d$ and messengers $X,
\bar X$ via the superpotential interactions \beq W = \lambda \Phi_1
H_u \bar X + \bar \lambda \Phi_2 H_d X \eeq and canonical kinetic
terms. The $X, \bar X$ are messengers of gauge mediation with the
quantum numbers of $H_u, H_d,$ and assume a messenger mass $M + F
\theta^2$ from supersymmetry breaking. In order to preserve
unification, we may readily assume that the $X$ and $\bar{X}$ are
part of complete $\mathbf{5} + \bar \mathbf{5}$ messenger multiplets
whose remaining components couple to the (heavy) Higgs
triplets.\footnote{We assume the GUT theory has a natural
doublet-triplet splitting mechanism. In this case, in addition to
the color triplets in the Higgs multiplets ($T, \bar T$) and the
messenger triplets ($X_T, \bar X_T$), we have auxiliary triplets
($T', \bar T'$) with the mass matrix in the basis ($T, \
T', \ X_T$) given by $$M_T = \left(\begin{array}{ccc} 0 & M_G & 0 \\
M_G & 0 & 0 \\ 0 & 0 & M \end{array}\right).$$  Such a missing
partner mechanism is natural in the context of orbifold GUTs. } We
will be interested in messenger masses corresponding to
intermediate-scale gauge mediation, with $M \simeq 10^7 - 10^8$ GeV.
In order to obtain weak-scale soft masses for the fields of the
Standard Model, this implies that the messenger fields experience
SUSY-breaking around $F \simeq 10^{13}$ GeV$^2$ with $\frac{F}{M}
\simeq 10^5$ GeV.  The gauge singlets $\Phi_1$ and $\Phi_2$ can be
thought of as moduli superfields whose only renormalizable couplings
to the Standard Model are those listed above.\footnote{Of course,
one might imagine the presence of many additional light
string-theoretic moduli in this theory with masses of order
$m_{3/2},$ subject to the usual cosmological constraints. However,
in the cosmology discussed here these light string moduli are
relatively uninteresting, and further use of ``moduli'' will
generally denote the singlets $\Phi_1, \Phi_2$ except where
otherwise noted}

Below the scale $M,$ we can integrate out the messengers to give
irrelevant interactions in the superpotential and K\"{a}hler
potential of the form \beq
 W &\supset& - \frac{\lambda \bar \lambda}{M} \Phi_1 \Phi_2 H_u H_d \\
 K &\supset& \frac{|\lambda|^2}{M^2} \Phi_1^\dag \Phi_1 H_u^\dag
 e^{-2 (g T_a V_a + \frac{1}{2} g' Y V_Y)} H_u + \frac{|\bar \lambda|^2}{M^2}
 \Phi_2^\dag \Phi_2 H_d^\dag e^{-2 (g T_a V_a + \frac{1}{2} g' Y/ V_Y)}H_d + ...
 \eeq
We are also interested in generating supersymmetric masses for the
$\Phi_1, \Phi_2$ and a $\mu$ term for the Higgses, which we can
assume to come from some Planck-suppressed operators in the
superpotential: \beq W \supset \frac{S^2}{M_P} \left( \Phi_1^2 +
\Phi_2^2 \right) + \frac{\bar{S}^2}{M_P} H_u H_d + Z X \bar{X}
\label{eq:flaton} \eeq Here $S, \bar{S}$ obtain vacuum expectation
values $\langle S \rangle \simeq \langle \bar{S} \rangle \simeq
10^{10}$ GeV to generate a $\mu$ term and supersymmetric mass terms
of order 100 GeV. $Z$ is a hidden-sector singlet whose vev $\langle
Z \rangle = M + F \theta^2$ breaks supersymmetry and imparts a mass
to the messenger fields.

\begin{figure}
   \centering
   \includegraphics[width=4in]{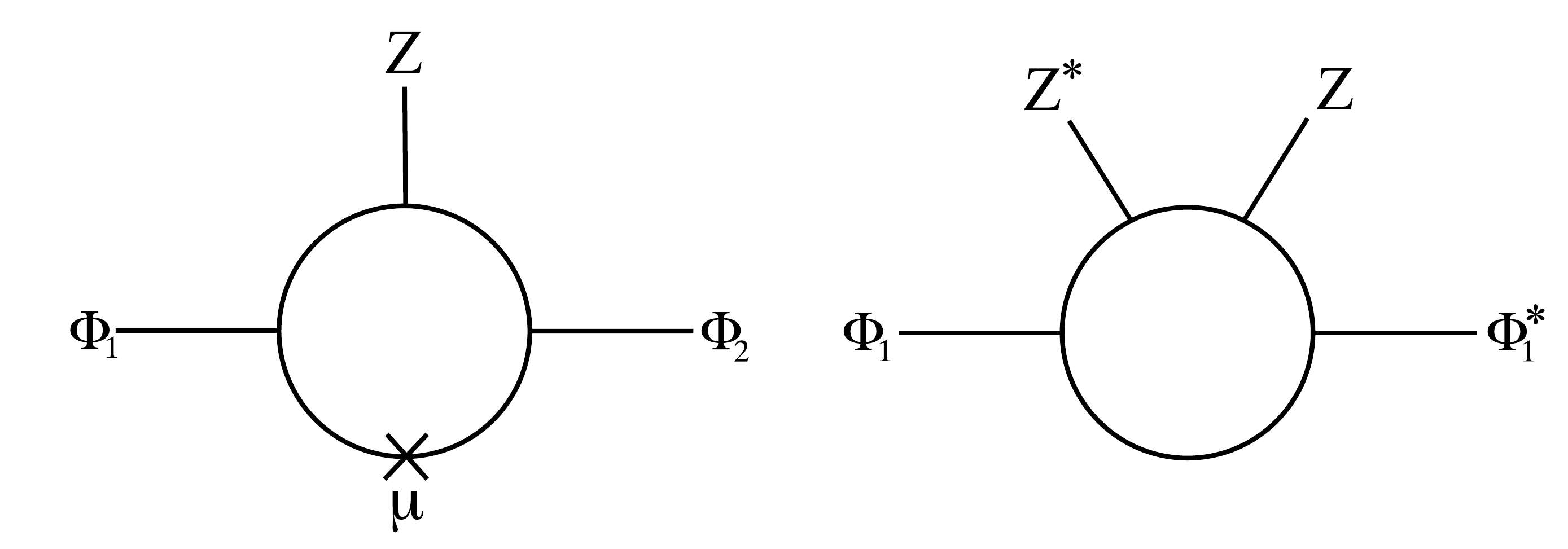}
   \caption{Diagrams generating a hard SUSY-breaking mass for $\psi_1  \psi_2$ (left) and soft scalar masses for $\phi_1^* \phi_1, \phi_2^* \phi_2$ (right). }
   \label{fig:diagram}
\end{figure}

There arise a number of additional contributions to the masses of
moduli and modulini when supersymmetry is broken by the $F$-term
expectation value of $Z.$ In addition to their supersymmetric masses
$m^2 \sim |\langle S \rangle|^4 / M_P^2,$ the moduli $\phi_1,
\phi_2$ obtain SUSY-breaking soft masses $m_{\phi_1}^2 \simeq
\frac{\lambda^2}{16 \pi^2} \frac{F^2}{M^2}, m_{\phi_2}^2 \simeq
\frac{\bar \lambda^2}{16 \pi^2} \frac{F^2}{M^2}$ of order 1 TeV (see
Fig.~\ref{fig:diagram}). Such large masses lead the moduli to decay
rapidly into modulini and MSSM fields well before BBN, rendering
them both cosmologically safe and uninteresting.

The mass spectrum of the modulini $\psi_1, \psi_2$ is somewhat more
interesting. The modulini obtain tree-level masses $m = \langle S^2
\rangle / M_P \simeq 10^2$ GeV from the VEV of $S.$ In
addition, a hard SUSY-breaking Dirac mass is generated radiatively
from the $\mu$ term when supersymmetry is broken (see
Fig.~\ref{fig:diagram}), and is of order $\delta m \simeq
\frac{\lambda \bar \lambda}{16 \pi^2} \frac{F}{M^2} \mu.$ The mass
eigenstates are then $\psi_{A,B} = \frac{1}{\sqrt{2}} (\psi_1 \pm
\psi_2)$ with masses $m_{A,B} = m \pm \delta m.$ It is precisely
this small splitting of the modulini mass eigenstates that will
prove interesting for the 511 keV line.

This theory possesses a number of amusing continuous and discrete
symmetries in addition to those of the Standard Model, including
both a $U(1)_{PQ}$ symmetry and a $\mathbb{Z}_4$ symmetry under
which $W \rightarrow - W.$ This $\mathbb{Z}_4$ symmetry includes the
familiar discrete $R$-parity that conventionally ensures the
stability of the MSSM LSP and guards against proton decay. The
appropriate charge assignments of the Higgs fields, messengers, and
gauge singlets are shown in Table I.  The $U(1)_{PQ}$ charges for
quarks and leptons are $-\frac{1}{2}$, while the $\mathbb{Z}_4$
quantum numbers are consistent with SU(5) with the $10 \ (\bar 5)$
with charge $i \ (-i)$. Among other things, the expectation value
$\langle \bar S \rangle$ breaks the PQ symmetry of the theory, and
generates an invisible axion with $f_a \simeq 10^{10}$ GeV.  It is
also worth noting that a tree-level Dirac mass for the $\Phi_{1,2}$
is forbidden by the symmetries of the theory.

\begin{table}[t]
\caption{Fields \& charge assignments.  Note the fields $\chi + \bar
\chi$ are a $\mathbf{5} + \bar \mathbf{5}$ of SU(5) while the fields
$X + \bar X$ are assumed to be the doublet members of a complete
$\mathbf{5} + \bar \mathbf{5}$.}
\begin{center}
\begin{tabular}{|c|c|c|c|c|} \hline
 & $SU(2)_L$ & $U(1)_Y$ & $U(1)_{PQ}$ & $\mathbb{Z}_4$ \\ \hline
 $H_u$ & $\square$ & $\frac{1}{2}$  & $1$    & $1$  \\
 $H_d$ & $\square$  & $-\frac{1}{2}$   & $1$  & $-1$  \\
 $X$      & $\square$  & $\frac{1}{2}$   & $0$  & $i$  \\
 $\bar{X}$& $\square$ & $-\frac{1}{2}$ & $0$  & $i$   \\
 $Z$& $\mathbf{1}$ & $0$ & $0$  & $1$   \\
 $\Phi_1$& $\mathbf{1}$ & $0$  & $-1$   & $i$   \\
 $\Phi_2$& $\mathbf{1}$ & $0$ & $-1$  & $-i$  \\
 $S$   & $\mathbf{1}$   & $0$   & $1$  & $-1$  \\
 $\bar{S}$&$\mathbf{1}$   & $0$   & $-1$  &  $-1$    \\
 $\chi$      & *  & *   & $\frac{1}{2}$  & $1$  \\
 $\bar{\chi}$& * & * & $\frac{1}{2}$  & $1$   \\   \hline
\end{tabular}
\end{center}
\label{default}
\end{table}%

The stability of the MSSM LSP is guaranteed by these symmetries. The
stability of the lightest modulino is guaranteed by a $\mathbb{Z}_2$
symmetry [moduli parity] where $\Phi_1, \Phi_2, X, \bar X$ are odd
and all other fields are even. If a cosmological abundance of this
modulino is generated in the early universe, it will remain as a
component of dark matter in the present era.

\section{Cosmology}

The thermal history of a theory with moduli coupled to the MSSM via
modestly irrelevant operators is rather interesting. Scattering with
the Higgs fields keeps the modulini in thermal equilibrium with the
Standard Model down to relatively low temperatures. The
cross-sections for Higgs-modulino scattering go like $\sigma v \sim
\frac{1}{M^2},$ significantly smaller than weak-scale
cross-sections. Modulini in thermal equilibrium freeze out when
$n_{eq} \langle \sigma v \rangle \sim H,$ which corresponds to a
temperature $T_f \simeq m_{\psi} / 8.$ Their freezout density is far
too high due to the relative inefficiency of annihilations, and at
freezout corresponds to a thermal abundance $\Omega_{\psi} h^2
\simeq 10^6.$  It is
therefore necessary to dilute any initial thermal abundance of
modulini. As is often the case for weakly-coupled fields with large
equilibrium number density, it is therefore reasonable to imagine a
scenario wherein the initial thermal abundance of modulini is
diluted by inflation. The universe may then reheat after inflation
to a temperature $T_R$ below the freezout temperature of the
modulini.

Of course, some thermal abundance of modulini will be produced by
thermal scattering even if the reheating temperature is too low to
produce modulini in equilibrium; the universe will be repopulated
with modulini produced by scattering and decay processes of Standard
Model fields in the thermal bath. It is therefore equally important
that reheating not produce an excessive thermal abundance from
scattering and decays of Standard Model particles.

It is relatively straightforward to compute the abundance of the
modulini due to thermal production. The time evolution of the
modulini number density $n_\psi$ (we need not distinguish between
$\psi_A$ and $\psi_B$ here, since their mass splittings are
sufficiently small to make their production rates virtually
identical)  is described by the Boltzmann equation, \beq \frac{d
n_{\psi}}{dt} + 3 H n_{\psi} \simeq 2 \langle \sigma(h h \rightarrow
\psi \psi) v \rangle n_h^2 + 2 \langle \Gamma(h \rightarrow \psi
\psi) \rangle n_h  - \text{inverse} \eeq Assuming small density
$n_\psi,$ we can make the approximation $3 H n_\psi \approx 0$ and
$n_{\psi}^{eq} \approx 0$. Then we have \beq \frac{d n_\psi}{dt}
\approx 2 \langle \sigma(h h \rightarrow \psi \psi) v \rangle n_h^2
+ 2 \langle \Gamma(h \rightarrow \psi \psi) \rangle n_h \eeq  A
thermal abundance of modulini is produced principally by Higgs
scattering and decays; the scattering cross-sections into modulini
are \beq
\sigma (h + h \rightarrow \psi_A + \psi_A) v= \frac{1}{256 \pi} \frac{s_\alpha^2 c_\alpha^2}{M^2} \left(1 - \frac{4 m_A^2}{s}\right)^{3/2} \\
\sigma (h + h \rightarrow \psi_B + \psi_B) v = \frac{1}{256 \pi} \frac{s_\alpha^2 c_\alpha^2}{M^2} \left(1 - \frac{4 m_B^2}{s}\right)^{3/2}
\eeq
Here $s_\alpha, c_\alpha$ denote
 $\sin, \cos$ of the angle determining the light neutral Higgs mass
 eigenstate, while $s_\beta, c_\beta$ are the $\sin, \cos$ of the angle
 $\beta$ with $\tan\beta = \frac{\langle H^0_u \rangle}{\langle H^0_d \rangle}$. For simplicity we have set $\lambda = \bar \lambda = 1,$ although the more general case is straightforward. Assuming that $m_h > 2 m_A, 2 m_B,$ the decay rates are given by
 \beq
 \Gamma(h \rightarrow \psi_A + \psi_A) = \frac{1}{256 \pi} (s_\alpha s_\beta - c_\alpha c_\beta)^2
 \frac{v^2}{M^2} m_h \left( 1 - \frac{4 m_A^2}{m_h^2} \right)^{3/2} \\
 \Gamma(h \rightarrow \psi_B + \psi_B ) = \frac{1}{256 \pi} (s_\alpha s_\beta - c_\alpha c_\beta)^2
 \frac{v^2}{M^2} m_h \left( 1 - \frac{4 m_B^2}{m_h^2} \right)^{3/2}
 \eeq
 Of course, if $m_h < 2 m_A,  2 m_B,$ the principle decays arise from other Higgs mass eigenstates such as the heavier CP even Higgs field $H;$ we will assume $m_h > 2 m_A, 2 m_B$ for simplicity.

In order to solve the Boltzmann equation, we may define the modulino
thermal production yield $Y_\psi = n_\psi /s,$ where $s$ is the
entropy density. The total yield is simply the sum of yields from
scattering and decay, $Y_\psi = Y^\sigma_{hh} + Y^\Gamma_h,$ where
\beq
Y^\sigma_{hh} &= \int_0^{T_R} dT \frac{2 \langle \sigma(hh \rightarrow \psi \psi) v \rangle n_h^2}{s H T} \\
Y^\Gamma_h &= \int_0^{T_R} dT \frac{2 \Gamma(h \rightarrow \psi \psi) n_{h}}{s H T}
\eeq

 {\it A posteriori}, we will be interested in reheating temperatures on the
 order of a few GeV, so we can assume $m_h \gg T_R.$ In this case the
 Higgs is essentially nonrelativistic, with number density well approximated by
  \beq
 n_h = 4 \left( \frac{m_h T}{2 \pi} \right)^{3/2} \exp[-m_h /T]
 \eeq

We also have $s = (2 \pi^2 / 45) g_{s \star} T^3$ and $H =
\frac{\pi}{\sqrt{90}} \sqrt{g_\star} \frac{T^2}{M_P}.$ Here $M_P$ is
the reduced Planck mass, $2.43 \times 10^{18}$ GeV. Having
determined the thermal yield $Y_\psi$, we may readily determine the
modulino relic abundance from thermal production via \beq
\Omega_{\psi} h^2 = m_\psi Y_\psi \frac{s(T_N)}{\rho_c / h^2}. \eeq

At reheating temperatures of a few GeV, scattering processes are
terribly inefficient -- both because thermal energies are well below threshold, and because of the $n_h^2$ Boltzmann suppression -- so that decay processes dominate over
annihilations. Nonetheless, production via decays is still
relatively effective and constrains $T_R \lesssim 10$ GeV in order
to avoid overproduction of modulini, as seen in Fig. 2.

\begin{figure}[t]
   \centering
      \includegraphics[width=4in]{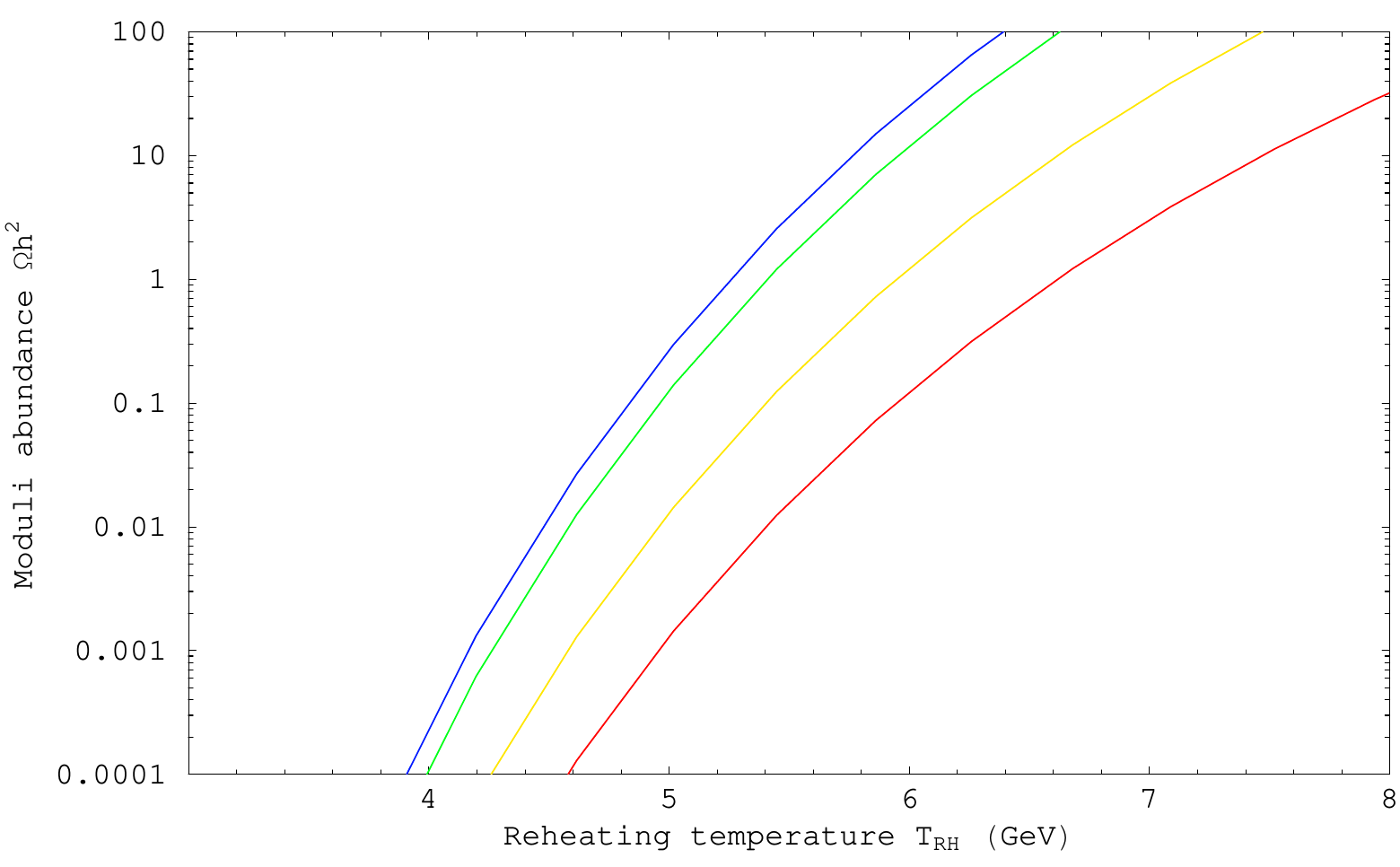}
   \caption{Relic abundance $\Omega_\psi h^2$ as a function of reheating temperature $T_R$. From blue to red, lines correspond to
   $m_\psi = 10^{-1}, 1, 10, 50$ GeV for $m_h = 150$ GeV and $M = 10^8$ GeV. }
   \label{fig:reheat}
\end{figure}

\subsection{Thermal inflation}
Supergravity theories generally suffer from light moduli with masses
of order the weak scale and long lifetimes. These light moduli tend
to dominate the energy density of the universe and decay after the
epoch of nucleosynthesis resulting in major cosmological problems
\cite{Coughlan:1983ci,Banks:1993en}.  Thermal inflation
\cite{Lyth:1995ka} is a mechanism introduced to solve the
cosmological moduli problem.   The universe is assumed to go through
an initial epoch of inflation followed by reheating thus solving the
homogeneity, isotropy and flatness problems and generating the
primordial density fluctuations. Then at a later time the universe
undergoes another period of thermal inflation with of order 9
e-foldings of expansion, hence diluting the light moduli.  The main
ingredient in thermal inflation is a ``flaton" field with mass $m$
of order $10^2$ to $10^3$ GeV and a potential whose minimum is
located at a scale $M_f$ of order $10^{10}$ to $10^{11}$ GeV. This
field induces a period of late time inflation with the flaton field
stuck at the origin in field space and dominating the energy density
of the universe starting at a temperature $T$ of order $10^6$ to
$10^7$ GeV. The inflation period then ends at the critical
temperature $T_C \sim m$, followed by a period of entropy production
ending with a reheat temperature $T_D \sim 1$ to $10^2$ GeV.  The
reheat temperature is safely above the epoch of nucleosynthesis.
However, with such a low reheat temperature one must necessarily be
concerned that the baryon number density has also been severely
diluted. Fortunately, it has been shown
\cite{Stewart:1996ai,Kim:2008yu} that a short period of Affleck-Dine
leptogenesis via preheating near $T_C$ can generate a sufficient
baryon number asymmetry.

It is remarkable that our model, for completely independent reasons,
contains all the ingredients required by
\cite{Stewart:1996ai,Kim:2008yu}.   In fact, we can interpret the
field $\bar S$ (Eqn. \ref{eq:flaton}) as the flaton.  It can couple
to new fields $\chi (= \mathbf{5}) + \bar \chi (= \bar \mathbf{5})$
that obtain mass of order $\langle \bar S \rangle = M_f$ via the
interaction \beq W \supset \bar S \chi \bar \chi .\eeq It is this
coupling which was shown to generate the negative mass squared for
the flaton at the origin. Note that the $PQ$ symmetry is also broken
at the scale $M_f = f_a$.

Finally at the reheat temperature $T_D$ we have already shown that
we obtain a satisfactory abundance of modulinos through thermal
processes.  These are then available to be a major component of the
dark matter of the universe and, as we now show, they can produce
the INTEGRAL signal.

\subsection{Dark Matter}

While the modulini $\psi_A, \psi_B$ may obtain a significant
cosmological abundance from reheating, it is by no means necessary
for them to constitute the sole components of dark matter. Indeed,
obtaining the measured dark matter relic abundance $\Omega h^2
\simeq 0.1$ for these modulini would involve a rather careful, and
perhaps unnatural, tuning of the reheating temperature. A smaller
abundance of modulini may still account for the observed 511 keV
line, while allowing a broader range of reheating temperatures. The
remaining dark matter relic abundance may be accounted for by
additional dark matter candidates. Indeed, our theory possesses an
abundance of such candidates. The gravitino mass in this theory may
range from $m_{3/2} \simeq 10 \text{ keV} - 100 \text{ MeV}$ and
comprise a component of dark matter as well.\footnote{The lower
bound arising if $F = F_0,$ i.e., the primordial SUSY-breaking scale
is the same as that felt by the gauge messengers. Of course, it is
entirely possible that $F \ll F_0,$ in which case the upper bound
comes from flavor constraints.} The low reheating temperature
following thermal inflation ensures that gravitinos are not
overproduced by thermal processes, and may instead have an abundance
close to the observed amount.

The theory also possesses an invisible axion and axino. The axino
mass arises through supersymmetry breaking, and is highly dependent
upon the details of the hidden sector. In models of gauge-mediated
SUSY-breaking with the gravitino mass range discussed above, the
axino mass may range from eV to GeV, and certainly may be lighter
than the gravitino \cite{Rajagopal:1990yx, Chun:1992zk}. In this
case, the axino is likely to be the MSSM LSP. Provided that it is
heavier than a few eV, the axino will be a warm or cold component of
dark matter. With such low reheat temperatures, the thermal
abundance of axinos is likely too small to account for a significant
fraction of dark matter \cite{Covi:2001nw}. However, for masses
above an MeV, the axinos may inherit a suitable freezout relic
abundance from the decay of heavier species such as the neutralino
\cite{Covi:1999ty}. In such a scenario, the axino may naturally
account for a significant fraction of the dark matter relic
abundance without fine-tuning the reheating temperature. Of course,
the axino may also be quite heavy in gauge mediation, leaving the
axion or various light string moduli as additional candidates
\cite{Carpenter:2009sw}.

The precise spectra and abundances of additional dark matter
candidates depends strongly on the details of the hidden sector and
PQ symmetry breaking, but there remain a plethora of additional
candidates that may obtain significant abundances from thermal
production or the freezout abundance of heavier species. Thus it is
not unreasonable for the modulini $\psi_A, \psi_B$ to constitute
part or all of the observed dark matter abundance without
prohibitive fine-tuning of the reheating temperature after thermal
inflation. Of course, whatever the dark matter candidate(s), the
prospects for direct detection are somewhat bleak; modulino, axino,
or gravitino dark matter are unlikely to produce a measurable
nuclear recoil signal in direct detection experiments.

\section{Decays of the modulino: INTEGRAL}

Thus far we have considered a model of gauge mediation incorporating
additional singlets $\Phi_1, \Phi_2$ that couple to the MSSM via
tree-level interactions with messengers and the Higgs fields $H_u,
H_d.$ The fermionic modulini of these singlets obtain both
supersymmetric and supersymmetry-breaking masses, which result in
splittings of $\mathcal{O}(MeV).$ The lightest of these modulini is
stable, and both modulini may obtain a sensible cosmological
abundance from thermal production after reheating. Due to the
smallness of their mass splittings, it is conceivable that the decay
of the heavier modulino into the lighter modulino may occur
sufficiently slowly to explain the 511 keV excess measured by
INTEGRAL.

Before we discuss the particulars of the decay process, it is worth
considering how the decays of dark matter might explain the observed
excess. The dark matter distribution in the galactic halo may be
parameterized as a function of $r$ by \beq \rho(r) \propto
\frac{1}{(r/a)^\gamma [1 + (r/a)^\gamma]^{(\beta - \gamma)/\alpha}}
\eeq for halo profile parameters $\alpha, \beta, \gamma;$ here $a$
is the distance from the galactic center where the power law breaks.
Near the galactic bulge, $\rho(r) \propto (r/a)^{-\gamma}.$
Following \cite{Hooper:2004qf}, the current SPI data with FWHM of $6^\circ$ is well fit by a cusped
profile with $\gamma \simeq 1.6,$ fairly consistent with results of
high-resolution N-body simulations \cite{Navarro:2003ew,
Diemand:2004wh}.

Normalizing the halo profile to the local dark matter density, one
thus finds, close to the galactic center, \beq \rho(r) \simeq
\frac{0.3 M_{\odot} / \text{pc}^3}{(r/1 \text{ kpc})^{1.6}} \eeq The
total mass within the $6^\circ$ circle of the INTEGRAL signal is thus \beq M_{INT}
= \int_0^{425 \, \text{pc}} \rho(r) 4 \pi r^2 dr \simeq 9.1 \times 10^{65}
\text{ GeV}. \eeq

The number of gamma rays contributing to the 511 keV line per
annihilated non-relativistic positron is given by $2(1-3 f/4),$
where $f = 0.967 \pm 0.022$ is the positronium fraction;
consequently, the decay rate producing positrons is $3.6$ times
larger than would be deduced from the gamma ray flux itself \cite{
Beacom:2005qv}. With this in mind, matching the rate of decays to
the observed 511 keV flux yields \beq \frac{M_{INT}}{m_{dm}
\tau_{dm}} \sim 3.6 \times \frac{1}{2} \Phi_{\text{exp}} 4 \pi
R_{GC}^2 \eeq where $R_{GC} \simeq 2.5 \times 10^{22} \text{ cm}$ is
the distance of the Earth from the galactic center. This suggests
the lifetime for dark matter decays in order to provide the INTEGRAL
signal is \beq \tau_{dm} \simeq 7 \times 10^{22} /
m_{dm}\text{(GeV)} \, \text{s}. \eeq Of course, it is not strictly
necessary for the decaying particle to comprise the entirety of dark
matter in the galaxy; for a decaying particle with abundance
$\Omega_\psi < \Omega_{dm},$ the lifetime required to explain
INTEGRAL is instead \beq \tau_{\psi} \simeq 7 \times 10^{22} \,
\left( \frac{\Omega_{\psi}}{\Omega_{dm}} \right) \, \frac{1}{
m_{\psi}\text{(GeV)}} \, \text{s}. \eeq provided that the
distribution of the decaying particle roughly tracks the dark matter
distribution discussed above.

The INTEGRAL signal also places rather significant constraints on
the injection energy of the positrons and electrons which produce
the observed gamma rays. Taking into account energy losses as well
as diffusion and delay, the annihilation of positrons produces gamma
rays at or below 511 keV. However, the emission of gamma rays by
positrons prior to annihilation may place a stringent constraint on
the injection energy. Firstly, the internal bremsstrahlung radiation
associated with positron production (a QED correction to the decay
process producing positrons) conflicts with COMPTEL and EGRET
diffuse gamma-ray constraints unless injection energies are
$\lesssim 20$ MeV \cite{Beacom:2004pe}. Secondly, the inflight
annihilation of relativistic positrons with electrons in the
interstellar medium -- a fate that may befall as many as 20\% of
energetic positrons -- places a much more stringent constraint of
$\lesssim 3$ MeV on the injection energy of individual positrons in
order to avoid exclusion by COMPTEL and EGRET bounds
\cite{Beacom:2005qv}.  It should be noted that this bound assumes
monoenergetic injection of positrons, and may be slightly relaxed by
the energy distribution of three-body phase space. Nonetheless, the
limits from internal bremsstrahlung and inflight annihilation
significantly proscribe the types of dark matter decays that could
explain the observed signal from INTEGRAL. Positron injection
energies of $\lesssim 3$ MeV from decaying dark matter allow two
possibilities: that the dark matter mass itself is
$\mathcal{O}(\text{MeV}),$ or that dark matter of mass $\gtrsim
\mathcal{O}(\text{MeV})$ is decaying into another state with a mass
difference $\delta m \sim \mathcal{O}(\text{MeV}).$ Our model is
suggestive of the latter possibility: a modulino of mass $m_{\psi}
\sim 100$ GeV (which may provide all or part of the dark matter in
our galaxy) decaying in a nearly-degenerate modulino with a mass
splitting of order $\sim$ few MeV.

The leading contribution to the decay $\psi_A \rightarrow \psi_B +
e^+ + e^-$ comes from the K\"{a}hler terms \beq K \supset
\frac{1}{M^2} \Phi_1^\dag \Phi_1 H_u^\dag e^{-2 (g T_a V_a +
\frac{1}{2} g' Y V_Y)} H_u + \frac{1}{M^2} \Phi_2^\dag \Phi_2
H_d^\dag e^{-2 (g T_a V_a + \frac{1}{2} g' Y V_Y)} H_d \eeq

\begin{figure}[t]
   \centering
 \includegraphics[width=4in]{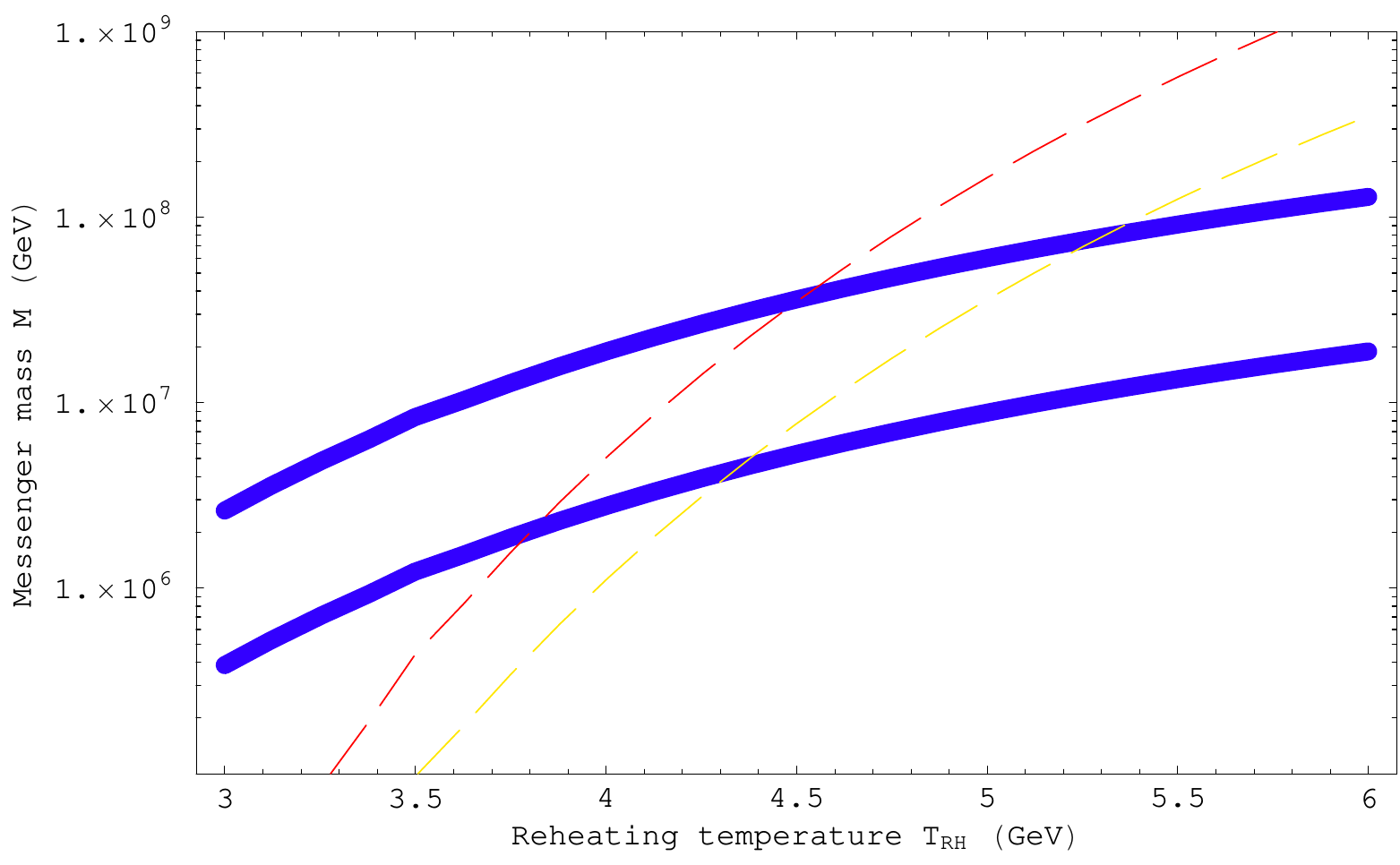}
   \caption{Messenger mass required to explain the INTEGRAL signal
   as a function of reheating temperature (and hence abundance).
   The region of interest lies between the two blue lines; the upper
   line corresponds to $\delta m = 10$ MeV, the lower to
   $\delta m = 1$ MeV. These bounds are only weak functions of the modulino mass; the width of each blue lines indicates the variation between $m_\psi = 1 - 50$ GeV. The red dashed line denotes $\Omega_\psi h^2 = 0.1$
   for $m_\psi = 50$ GeV (and $\Omega_\psi h^2 < 0.1$ to
   the left of the line), while the orange dashed line denotes $\Omega_\psi h^2 = 0.1$ for $m_\psi = 1$ GeV. }
   \label{fig:mass}
\end{figure}

These K\"{a}hler terms result in interaction terms between the
modulini and the $Z$ boson which, in the mass eigenbasis of the
modulini, take the form \beq V \supset  \frac{M_Z v}{4 M^2} \left[
(s_\beta^2 - c_\beta^2) \psi_A \sigma^\mu \psi_A^\dag Z_\mu +
(s_\beta^2 - c_\beta^2) \psi_B \sigma^\mu \psi_B^\dag Z_\mu
 +\psi_A \sigma^\mu \psi_B^\dag Z_\mu +  \psi_B \sigma^\mu \psi_A^\dag Z_\mu \right]
\eeq The smallness of the splitting between $\psi_A$ and $\psi_B$
constrains the $\psi_A$ to decay via an off-shell $Z$ into electrons
and positrons. The decay rate for this process is given by \beq
\Gamma(\psi_A \rightarrow \psi_B + e^+ + e^-)=\frac{1}{240 \pi^2}
\frac{\alpha  v^2}{ s_w^2  m_W^2}  \left[ \left(-\half + s_w^2
\right)^2 + s_w^4 \right] \frac{(\delta m)^5}{M^4} \eeq Using
$\alpha(m_Z) \approx 1/128, s_w^2 \approx 0.231, m_W \approx 80.4$
GeV, $v= 174$ GeV, the lifetime of the $\psi_A$ is simply \beq
\tau_{\psi_A} \simeq 7.83 \times 10^{-20} \frac{M^4}{(\delta m)^5}
\text{ s} \eeq  For $m_{\psi_A} \simeq 50$ GeV and messenger scale
$M \simeq 10^8$ GeV, the INTEGRAL signal may be matched for $\delta
m \approx 20$ MeV, assuming dark matter relic abundance for the
$\psi_A.$ Slightly smaller values of $M$ readily accommodate splittings $\delta m = 1 - 10$ MeV consistent with observational constraints. Of course, it is not necessary for the modulini to
comprise the entirety of the dark matter relic abundance; moreover
the INTEGRAL signal may be produced for a range of $m_{\psi_A},
\delta m, M,$ and $\Omega_\psi h^2$ (see Fig.~\ref{fig:mass}).

One might also be interested in decays via superpotential terms such
as \beq W \supset -\frac{\lambda \bar \lambda}{M} \Phi_1 \Phi_2 H_u
H_d \eeq which appears to generate decays into $e^+ e^-$ via Higgs
exchange. However, this term results in an effective interaction of
the form \beq \mathcal{L} \supset \frac{1}{4} \frac{\lambda \bar
\lambda}{M} \frac{m_e}{m_h^2} (s_\alpha^2 \tan \beta + s_\alpha
c_\alpha) \psi_1  \psi_2 \bar e e . \eeq  For the
mass spectrum considered here, where the mass eigenstates $\psi_A,
\psi_B$ are maximally-mixed combinations of the fields $\psi_1,
\psi_2,$ this interaction does not lead to a decay between the two
mass eigenstates.  For more general theories where this is not the
case, the decay mediated via the Higgs results in a coincidentally
similar rate to that via $Z$ bosons considered above, despite the
different parametric dependence on $M.$

\section{Diffuse gamma ray production}

Any theory giving rise to the 511 keV signal via dark matter decays
or annihilations inevitably makes an additional prediction:
potentially significant contributions to the isotropic diffuse
photon background [iDPB] due to internal bremsstrahlung from the
final-state electrons and positrons. Such an isotropic, diffuse
gamma ray signal coming from the galactic halo would be a suggestive
signal of dark matter decays.\footnote{As opposed to dark matter
annihilation which would be confined predominantly to the galactic
center.}

The isotropic gamma-ray background has been measured at high
galactic latitudes ($\gtrsim 10^\circ$) by COMPTEL (the Compton
Imaging Telescope) and SMM (the Solar Maximum Mission) over ranges
$0.8-30$ MeV \cite{Weidenspointner} and $0.3-7$ MeV \cite{Watanabe}
(respectively), while INTEGRAL has measured a diffuse photon
background over the range $5-100$ keV \cite{Churazov:2006bk}. The
analysis of iDPB in the MeV region is made difficult by both
instrumental and cosmic ray backgrounds and requires careful
analysis. While the iDPB is well-explained at energies below a few
hundred keV (where active galactic nuclei dominate) and above 10 MeV
(where blazars may account for the measured signal), there is
currently no known astrophysical source capable of accounting for
the entirety of the observed iDPB between 1 and 5 MeV.\footnote{Type
Ia supernovae do contribute gamma rays below 5 MeV, but the most
recent astrophysical data suggest that they cannot account for the
entire spectrum \cite{Strigari:2005hu}.} If anything, the observed
iDPB between 1 and 5 MeV may be indicative of new physics such as
the decay or annihilation of dark matter \cite{Cembranos:2008bw}.

The internal bremsstrahlung signal from dark matter decays in the
galactic halo may be readily estimated from \cite{Beacom:2004pe},
and falls well within the observed iDPB signal in the MeV range
\cite{Yuksel:2007dr}. However, future measurements of the isotropic
gamma ray background, coupled with an improved understanding of
astrophysical sources in the MeV range, may significantly improve
these bounds. The detection of diffuse gamma rays from internal
bremsstrahlung in the galactic halo would certainly be a suggestive
indication of dark matter decays associated with the 511 keV line,
and is an unambiguous additional prediction of this model.

\section{Conclusions}

We have considered a relatively minimal extension of the MSSM with
gauge mediated supersymmetry breaking, wherein additional MSSM gauge
singlet modulinos may account for some or all of the observed dark
matter. Natural SUSY breaking-induced mass splittings between the
modulini are of the appropriate scale to provide both the decay rate
and positron injection energy required to explain the 511 keV
signal. The cosmology of such a model is no different from that
required to solve the gravitino and moduli problems of
gauge-mediated supersymmetry breaking, and coincidentally produces a
suitable cosmological abundance of the modulini. While the lightest
modulino is stable, the modulini need not comprise the entire dark
matter of the galaxy; gravitinos or axinos may also constitute a
sizable fraction of dark matter without undue tuning of the
reheating temperature. Whatever the exact composition of dark
matter, the prospects for direct detection are invariably bleak; the
interactions of the modulini and axino are far too small to be
measured above irreducible backgrounds in nuclear recoil
experiments. Additional signatures of this scenario must come from
the sky; internal bremsstrahlung from the decay of the heavier
modulino inevitably produces diffuse gamma rays in the galactic halo
that may be measured by future experiments.

\acknowledgments

We would like to thank John Beacom and Matt Kistler for useful
discussions. NJC would like to acknowledge the hospitality and support of the
Department of Physics at The Ohio State University and the Dalitz
Institute at Oxford University, where parts of this work were
completed. SR would like to acknowledge the hospitality and support of the Stanford Institute for Theoretical Physics and the Dalitz Institute at Oxford University. 
NJC is supported in part by the NSF GRFP, the NSF under
contract PHY-9870115, and the Stanford Institute for Theoretical
Physics. SR is supported in part by DOE grant DOE/ER/01545-882.

\bibliography{modulino}
\end{document}